\documentclass{amsart}   
\usepackage{todonotes}

\pdfoutput=1

\usepackage{amssymb}
\usepackage{amsmath}
\usepackage[utf8]{inputenc}
\usepackage{multicol}
\usepackage{placeins}
\usepackage{float}
\usepackage{epsf}
\usepackage{longtable}
\usepackage{multirow}
\def\mb{\mathbf}

\def\laa{\leftarrow}

\def\o{\overline}

\def\le{\leqslant}
 \def\ge{\geqslant}
\def\ra{\rightarrow}
\usepackage{algorithm}
\usepackage{algpseudocode}
 
\usepackage{lipsum}
\usepackage{amsfonts}
\usepackage{graphicx}
\usepackage{epstopdf}
\ifpdf
  \DeclareGraphicsExtensions{.eps,.pdf,.png,.jpg}
\else
  \DeclareGraphicsExtensions{.eps}
\fi

\usepackage{amsopn}
 
 \theoremstyle{plain}
\newtheorem{theorem}{Theorem}[section]
\newtheorem{lemma}[theorem]{Lemma}
\newtheorem{proposition}[theorem]{Proposition}
\newtheorem{corollary}[theorem]{Corollary}

\theoremstyle{definition}
\newtheorem{definition}[theorem]{Definition}
\newtheorem{example}[theorem]{Example}
\newtheorem{remark}[theorem]{Remark}

\newenvironment{case}[1][\unskip]{\par\medskip\noindent\textbf{Case #1:}\noindent}{}

\begin{document}
\title{Minor complexities  of finite operations} 

\author{Slavcho Shtrakov}
\email{shtrakov@swu.bg} \urladdr{http://shtrakov.swu.bg/}
\address{Department of Computer Science, South-West University, Blagoevgrad} 

 \setlength\headheight{12pt}
  
\date{}

\begin{abstract}
In this paper we present a new class of complexity measures, induced by a new data structure for representing $k$-valued functions (operations), called minor decision diagram. The  results are presented  in terms of Multi-Valued Logic circuits (MVL-circuits), ordered decision diagrams, formulas and minor decomposition trees.  
   When assigning values to some variables in a function $f$ the resulting
function is a  subfunction of $f$, and when identifying some variables  the 
resulting function is a  minor of $f$.  
A set $M$ of essential variables in $f$ is separable if there is a  subfunction of $f$, whose set of essential variables is $M$.
The essential arity gap 
$gap(f)$ of the function $f$ is  the minimum number of essential variables in $f$ which become fictive when identifying  distinct essential variables in $f$. We prove that, if a function $f$ has non-trivial arity gap ($gap(f)\ge  2$), then all sets of essential variables in $f$ are separable.
We define  equivalence relations which classify the functions of $k$-valued 
logic
into classes with the same minor complexities. These relations induce  
transformation groups
which are compared with the  subgroups of the restricted affine group
(RAG) and the groups determined by the equivalence relations with respect to the subfunctions, implementations 
and separable sets in functions. These methods provide a detailed classification of $n$-ary $k$-valued functions for small values of $n$ and $k$.
\end{abstract}

 \maketitle
\section{Introduction}\label{sec1}

The complexity of finite operations is still one of the
fundamental tasks
in the theory of computation and besides classical methods like
substitution
or degree arguments a bunch of combinatorial, and algebraic techniques have been
introduced to tackle this extremely difficult problem.

A \emph{logic gate} is a physical device that realizes a Boolean function. A \emph{logic circuit} is a directed acyclic graph in which all vertices except input   vertices carry the labels of gates. When realizing functions are taken from the $k$-valued logic the circuit is called the \emph{ $(k,n)$-circuit or Multi-Valued Logic  circuit (MVL-circuit)}.

To  move  from logical circuits to MVL-circuits, researchers  attempt to adapt CMOS (complementary metal oxide semiconductor), I$^2$L (integrated injection logic) and ECL (emitter-coupled logic) technologies to implement the many-valued and fuzzy logics gates. The MVL-circuits offer more potential opportunities for the improvement of present VLSI circuit designs. For instance, MVL-circuits are well-applied  in  memory technology as flash memory, dynamic RAM, and in algebraic circuits \cite{dub2}.

Computational complexity is examined in concrete and abstract terms. The concrete analysis is based on  models that capture the exchange of space for time. It  is also performed via   the knowledge about  circuit complexity of functions. The abstract analysis is done via complexity classes, the classification of data structures, functions etc. by the time and/or space they need.

There are two key methods for reduction (computing) of the finite functions which are realized by assigning constants or variables  to their inputs.   Then  the resulting objects are:  subfunctions or minors, respectively. 
These reductions are also naturally suited to  complexity measures, which illustrate "difficulty" of  computing  as the number of subfunctions, implementations, and minors of the functions.

Another   topic in   Complexity Theory is to classify  finite functions by their complexity   such that the functions are grouped into equivalence classes with same evaluations of the corresponding complexities. Each equivalence relation in the algebra $P_k^n$ of $k$-valued functions determines a transformation group  whose orbits are the equivalence classes. Using the lattice of Restricted Affine Groups (RAG) in \cite{sh5} we have obtained  upper bounds of different combinatorial parameters of several natural equivalences in $P_k^n$  for small values of $k$ and $n$. 
In the present paper we follow this line to study assigning (not necessarily unique) variable  names to some of the input variables in a function $f$.  This method of computing consists of equalizing the values of several inputs of $f$.

 Section \ref{sec2}  introduces the basic definitions and notation of 
separable sets, subfunctions, minors, arity gap, etc. An important result, namely  if a function has non-trivial arity gap then all its sets of essential variables are separable, complements this section.
Section \ref{sec3} examines the minor
decomposition trees (MDTs),  minor decision diagrams (MDDs), and
minor complexities of discrete functions.
Equivalence relations and transformation groups 
 concerning the number of   minors in functions is the topic of  Section \ref{sec31}. 
Classification of Boolean (switching) functions of a "small" number of their
essential variables is presented in Section \ref{sec4}. The  Appendix examines an algorithm for counting the minor complexities of functions and provides a full classification of all ternary Boolean functions by these complexities.

\section{Subfunctions and minors of functions}\label{sec2}

A  \emph{discrete function} $f$ is defined as a mapping: $f:A\to B$ where the domain 
$A={\times}_{i=1}^n A_i$   and the range
 $B$ are 
non-empty finite or countable sets. 
   Let $X=\{x_1,x_2,\ldots \}$ be a countable set of variables and let 
$X_n=\{x_1,x_2,\ldots,x_n\}$  denote the set of the first $n$ variables in $X$. Let $k$ be a natural 
number
with $k\ge 2$. Let $Z_k$ denote the set $Z_k=\{0,1,\ldots,k-1\}$. The operations  addition  $"\oplus"$  and product $"."$ modulo $k$   constitute $Z_k$ as a ring. An \emph{$n$-ary $k$-valued   function
(operation) on $Z_k$ } is a mapping 
$f: Z_k^n\to Z_k$ for some natural
number $n$, called \emph{the arity}   of $f$.  $P_k^n$ denotes the set of all $n$-ary 
$k$-valued
functions and
$P_k=\bigcup_{n=1}^\infty P_k^n$  
is called \emph{the algebra of $k$-valued logic}. It is well-known fact that there are $k^{k^n}$ functions in $P_k^n$.
For simplicity, let us assume that throughout the paper we shall consider $k$-valued functions, only. 

For a given  variable $x$ and $\alpha\in Z_k$, $x^\alpha$ is defined as follows:
\[
  x^\alpha=\left\{\begin{array}{ccc}
             1 \  &\  if \  &\  x=\alpha \\
             0 & if & x\neq\alpha.
           \end{array}
           \right.
\]
The ring-sum expansion (RSE) of a function $f$ is the sum modulo $k$ of a constant
and products of variables $x_i$ or $x_i^\alpha$, (for $\alpha$, $\alpha\in Z_k$) of $f$. For example, $1\oplus x_1x_2^2$  is a RSE of the function $f$ in the algebra $P_3^2$, with $f(1,2)=2$, $f(2,2)=0$ and $f=1$, otherwise.
 Any $k$ instances of the same product in the RSE can be eliminated since they
sum to $0$. 
   Throughout  the present paper, we shall use RSE-representation of functions.

  Let $f\in P_k^n$ and let  $var(f)=\{x_1,\ldots,x_n\}$  be the set of all 
variables, which occur in $f$. 
   We say that the $i$-th variable $x_i\in var(f)$ is  \emph{essential} 
  in $f$, or $f$  \emph{
essentially depends} on $x_i$, if there exist values
$a_1,\ldots,a_n,b\in Z_k$, such that
\[f(a_1,\ldots,a_{i-1},a_{i},a_{i+1},\ldots,a_n)\neq  
f(a_1,\ldots,a_{i-1},b,a_{i+1},\ldots,a_n).\]
 
The set of all essential variables in the function $f$ is denoted  
$Ess(f)$ and
$ess(f)=|Ess(f)|$. The variables from $var(f)$ which
are not essential in
 $f\in P_k^n$ are called \emph{inessential} or \emph{fictive}.

Let $x_i$  be an essential variable in $f$ and let $c$ be a constant from $Z_k$. 
The
function $g=f(x_i=c)$ obtained from $f\in P_{k}^{n}$ by assigning the constant  $c$ to the variable 
$x_i$   is
called a \emph{simple subfunction of $f$} (sometimes termed a \emph{cofactor} or a \emph{restriction}).    
 When $g$ is a simple subfunction of $f$ we  write $f\succ g$. The 
transitive closure of $\succ$ is denoted  $\succeq$.   $Sub(f)=\{g\ | \ 
f\succeq g\}$ is the set of all subfunctions of $f$ and $sub(f)=|Sub(f)|$.

We say that each subfunction $g$ of $f$ is a reduction to $f$ via the \emph{subfunction relationship}.

A non-empty set $M$ of essential variables in the function $f$
 is called \emph{separable} in $f$ if there exists a subfunction $g$,
$f\succeq g$ such that $M=Ess(g)$.
   $Sep(f)$ denotes the set of  all the separable sets in $f$ and
$sep(f)=|Sep(f)|$.
 
 An essential variable $x_i$ in a function $f\in P_k^n$ is called a \emph{strongly essential variable} in $f$ if there is a constant $c_i$ such that $Ess(f(x_i=c_i)=Ess(f)\setminus \{x_i\}$. The set of all strongly essential variables in $f$ is denoted  $SEss(f)$.

The following lemma is independently proved by  K. Chimev \cite{ch51} and A. Salomaa \cite{sal} in different variations.
\begin{lemma}\label{l2} \cite{ch51}
Let $f$ be a  function. If $ess(f)\ge 2$ then $f$ has at least two strongly essential variables, i.e. $|SEss(f)|\ge 2$.
\end{lemma} 

Let $x_i$ and  $x_j$ be two distinct essential variables in  $f$. The
function $h$ is obtained from $f\in P_{k}^{n}$ by \emph{identifying (collapsing) the variables $x_i$ and $x_j$}, if
\[h(a_1,\ldots,a_{i-1},a_i,a_{i+1},\ldots,a_n)=
f(a_1,\ldots,a_{i-1},a_j,a_{i+1},
\ldots,a_n),\]
for all $(a_1,\ldots,a_n)\in Z_k^n$.

Briefly, when  $h$ is obtained from $f,$ by identifying the
variable $x_i$ with $x_j$, we  write $h=f_{i\leftarrow j}$ and $h$ is
called \emph{a simple identification minor of $f$} \cite{mig2}. Clearly, $ess(f_{i\leftarrow j})< ess(f)$, because
$x_i\notin Ess(f_{i\leftarrow j})$, but it has to be essential in
$f$.
When $h$ is a simple identification minor of $f$ we  write $f\rhd h$. The 
transitive closure of $\rhd$ is denoted   $\unrhd$. 
$Mnr(f)=\{h\ |\ f\unrhd h\}$
is the set of all distinct minors of $f$ and $mnr(f)=|Mnr(f)|$.
Let $h$, $f\unrhd h$ be an identification minor of $f$. The natural number $r=ess(f)-ess(h)$, $r\ge 1$ is called the \emph{order} of the minor $h$ of $f$.

We say that each minor $h$ of $f$ is a reduction to $f$ via the \emph{minor relationship}.

 Let $Mnr_m(f)$ denote the set $Mnr_m(f)=\{g\ |\ g\in Mnr(f)\ \&\ ess(g)=m\}$ and  let $mnr_m(f)=|Mnr_m(f)|$, for all $m$, $m\le n-1$.

Let  $f\in P_k^n$ be an $n$-ary $k$-valued function. The   \emph{essential 
arity gap} (shortly
\emph{arity gap} or \emph{gap}) of $f$ is defined as follows
\[gap(f)=ess(f)-\max_{h\in Mnr(f)}ess(h).\]

Let $2\le p\le m$. We let $G_{p,k}^m$ denote the set of all $k$-valued 
functions
which essentially  depend on $m$ variables whose arity gap is equal to
$p$, i.e.
\[G_{p,k}^m=\{f\in P_k^n\ |\ ess(f)=m\ \&\ gap(f)=p\}.\]

 We say that the arity gap of $f$ is \emph{non-trivial}  if $gap(f)\ge 2$. 
 It is natural to expect  that the functions with "huge" gap, have to be
more simple for realization by MVL-circuits and functional
schemas when computing by identifying  variables. For more results about the arity gap we refer \cite{ch51,mig2,sal,sht_1,sht_2,sht_3,ros}.

\begin{definition}\label{d1}
Two functions $g$ and $h$   are called \emph{equivalent (non-distinct as mappings)} (written $g\equiv h$) if $g$ can be obtained from $h$ by permutation of variables, introduction or deletion of inessential variables.
 \end{definition}

As mentioned earlier, there are two general ways for reduction of functions - by subfunctions or by  minors. The complexities of these processes we call the \emph{subfunction or minor complexities}, respectively.

\begin{figure}[hbt]
\centering
\includegraphics[trim = 0 0 0 0]{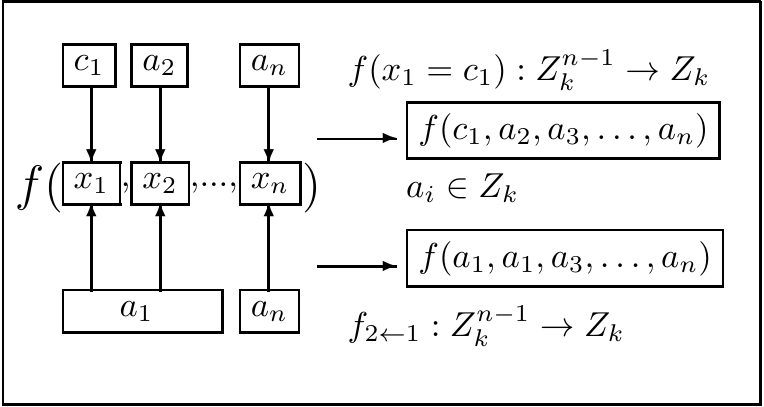}
\caption{Simple subfunction and simple minor of functions}\label{f7}
\end{figure}  
 
Figure \ref{f7} illustrates the difference between these two models of computing. The reduction to simple subfunction $f(x_1=c_1)$ is presented in the top    of the figure, and computing the identification minor $f_{2\laa 1}$ is shown  in the bottom of the figure. The resulting  functions of the both computations have the same domain - $Z_k^{n-1}$ and the same range - $Z_k$ (see the right side of Figure \ref{f7}). In practice,   both resulting functions  can be seen as partial mappings of $Z_k^n$ into $Z_k$. The subfunction  $f(x_1=c_1)$ is obtained by fixing the first input ($x_1$), whereas  the minor $f_{2\laa 1}$  has   two  identical inputs (the left-lower corner of the figure). 
Another obvious difference between these concepts is the following: Each identification minor can be decomposed into subfunctions, but there are subfunctions which can not be decomposed into minors. For example, let  $f=x_1\oplus x_2\oplus x_3$ be a Boolean function.  It is easy to see that the subfunction $f(x_1=1)=x_2\oplus x_3\oplus 1$ can not be decomposed into any minors of  $f$.


Many computations,
constructions, processes, translations, mappings and so on, can be modeled
as stepwise transformations of objects known as reduction systems. 
\emph{Abstract Reduction Systems (ARS)} play an important role in various areas such as abstract data
type specification, functional programming, automated deductions, etc. \cite{klo,sht_4}  The concepts and properties of ARS also apply to other rewrite
systems such  as string rewrite systems (Thue systems),  tree rewrite systems, graph
grammars, etc. For more
detailed facts about ARS we refer to J. W. Klop and Roel de Vrijer
\cite{klo}.
An  {ARS} in $P_k^n$ is a structure $  W=\langle P_k^n, \{\rightarrow_i\}_{i\in I},
\rangle$, where $\{\rightarrow_i\}_{i\in I}$ is a family of binary relations on
$P_k^n$, called \emph{reductions or rewrite relations}.
For a reduction $\rightarrow_i$ the transitive and reflexive closure is denoted
  $\twoheadrightarrow_i$.
A function $g\in P_k^n$ is a \emph{normal form} if there is no $h\in P_k^n$
such that $g\rightarrow_i h$.
  In all  
different branches of rewriting two basic concepts occur,
 known as termination (guaranteeing the
existence
of normal forms) and confluence (securing the uniqueness of normal
forms). 

A reduction $\rightarrow_i$ has the \emph{unique normal form property} (UN) if whenever
$t,r\in P_k^n$ are normal forms obtained by applying the reductions  $\rightarrow_i$ on a function $f\in P_k^n$ then $t$ and $r$ are equivalent (non-distinct as mappings).

The computations on functions proposed in the present paper can be regarded as an ARS, namely: $W=
\langle P_k^n,
\{\succ, \rhd\}\rangle$.
Next, we show that  $\rhd$ completes the reduction process with unique normal form, whereas $\succ$ has not unique normal form property.

A reduction $\rightarrow$ is \emph{terminating} (or \emph{strongly normalizing} SN) if every reduction sequence
$f\rightarrow f_1\rightarrow f_2\dots$ eventually must terminate.
A reduction $\rightarrow$ is \emph{weakly
confluent} (or   \emph{has weakly Church-Rosser property} WCR) if $f\rightarrow r$ and $f\rightarrow v$ imply
that there is  $w\in P_k^n$ such that $r\twoheadrightarrow w$ and $v\twoheadrightarrow
w$.

\begin{theorem}\label{t3}~~

\begin{enumerate}
\item[(i)] The reduction $\rhd$  is UN;
\item[(ii)] The reduction  $\succ$ is not WCR, but it is SN.
\end{enumerate}
\end{theorem}
\begin{proof} (i) 
(SN) Clearly, if $f\rhd g$ then  $ess(f)>ess(g)$.
Since the number of essential variables  $ess(f_i)$  of the functions $f_i$ in any
reduction sequence $f\rhd f_1\rhd\ldots\rhd f_i\rhd\ldots$  strongly decrease, it follows that the sequence eventually must
terminate, i.e.  the reduction is terminating.

(WCR) Let $f$ be a function  and $f\rhd g$, and $f\rhd h$.
Let $t$ and $r$ be normal forms such that $g\unrhd t$ and $h\unrhd r$. Note that each normal form is a resulting minor obtained by  collapsing  all the essential variables  in $f$. Hence, $ess(t)\le 1$ and $ess(r)\le 1$.
 Then we have 
$t=f(x_j,\ldots,x_j)$, 
for some $x_j\in Ess(f)$ and $r=f(x_i,\ldots,x_i)$, 
for some $x_i\in Ess(f)$, and hence, 
\[t=f(x_j,\ldots,x_j)\equiv f(x_i,\ldots,x_i)=r.\]
 
Now, (i)
  follows from Newman's Lemma (Theorem 1.2.1. \cite{klo}), which states that  WCR \& SN $\Rightarrow$ UN. \qed
  
  (ii) is clear.
\end{proof}

Below, we also establish a key theorem which states that the functions with simplest minor complexity (with non-trivial arity gap) have extremely complex representations with respect to  the number of their subfunctions, and separable sets.

\begin{lemma}\label{t1.6}
Let $N\in Sep(f)$. If there exist $m$ constants $c_1,\ldots,c_m\in Z_k$ such that  $N\cap Ess(g_i)=\emptyset$ where $g_i=f(x_i=c_i)$ 
for 
$1\le i\le m$   then $M\cup N\in Sep(f)$ for all 
$M\neq\emptyset$,  $M\subseteq \{x_1,\ldots,x_m\}$.
\end{lemma}
\begin{proof} It suffices to look only at the set $M= \{x_1,\ldots,x_m\}$. First, assume that $M\cap N=\emptyset$ and without loss of 
generality let us assume
 $N=\{x_{m+1},\ldots,x_s\}$, $m< s\le n$. Since  
  $N\in Sep(f)$, there exists a vector of constants, say
$\o{d}=(d_{s+1},\ldots,d_n)\in Z_k^{n-s}$ 
such that $N\subseteq Ess(g)$, where
 \[g=f(x_{s+1}=d_{s+1},\ldots,x_n=d_n).\] 

Let us   fix an arbitrary variable from $N$, say the variable  $x_s\in N$. 
Then there exist $s-m-1$ constants $d_{m+1},\ldots,d_{s-1}\in Z_k$ such that $x_s\in 
Ess(h)$ where 
\[h=g(x_{m+1}=d_{m+1},\ldots,x_{s-1}=d_{s-1}).\]
 We have to prove that 
$M\subseteq Ess(h)$. Let us suppose the opposite, i.e. there is a 
variable, say $x_1\in M$ which is inessential in $h$. Since $x_1\in  
\{x_1,\ldots,x_m\}$, there is a value $c_1\in Z_k$ such that $N\cap 
Ess(t)=\emptyset$ where 
$t=f(x_1=c_1)$.
Our supposition shows that 
$h=h(x_1=c_1)$ and hence, $N\cap Ess(h)=\emptyset$, i.e. $x_s\notin Ess(h)$, which is a 
contradiction. 
Consequently, $M=Ess(h)$. Then $g\succeq h$ implies 
$M\subseteq Ess(g)$ and hence, $M\cup N=Ess(g)$ which establishes  that $M\cup N\in 
Sep(f)$.

Second, let $M\cap N\neq\emptyset$. Then we can pick $P=M\setminus N$ and hence, 
$P\subseteq  \{x_1,\ldots,x_m\}$, $P\cap N=\emptyset$, and $N\in Sep(f)$. As  shown, above $P\cup N\in Sep(f)$ and $M\cup 
N\in Sep(f)$, as desired.\qed
\end{proof}
\begin{corollary}\label{c3}
Let $x_i$ and $x_j$ be two distinct essential variables in $f$. If there is a constant $c$, $c\in Z_k$ such that $f(x_i=c)$ does not essentially depend on $x_j$ then $\{x_i,x_j\}\in Sep(f)$.
\end{corollary}
Next, we turn our attention to   relationship between essential arity gap and separable sets in functions.
\begin{theorem}\label{t2} 
Let $f\in P_k^n$. If
  $gap(f)\ge 2$  then all   non-empty sets of essential variables   are separable in $f$.
 \end{theorem}
 \begin{proof} Without loss of generality, let us assume that $Ess(f)=\{x_1,\ldots,x_n\}$.
  Let $M$ be an arbitrary non-empty set of essential variables in $f$.  We shall prove that $M\in Sep(f)$ by considering cases.
The theorem is given to be true if $n\le 2$. Next, we assume $n>2$.
\begin{case}[1] 
$gap(f)=2$, $n\ge 3$ and $k=2$.

If $n\le 4$  then we are done because of Theorem 3.2 and Theorem 3.3 in \cite{sht_1}.

Let $n\ge 5$. From Theorem 3.4 in \cite{sht_1} it follows that 
\[ f= \bigoplus_{\alpha_1\oplus\ldots\oplus\alpha_n=1}x_1^{\alpha_1}\ldots
 x_n^{\alpha_n}\quad\mbox{or}\quad
 f= \bigoplus_{\alpha_1\oplus\ldots\oplus\alpha_n=0}x_1^{\alpha_1}\ldots
 x_n^{\alpha_n}.\]
  Clearly, $Ess(f)=X_n$. Suppose, with no loss of generality that $M=\{x_1,\ldots,x_m\}$, $m<n$ and 
 \[f= \bigoplus_{\alpha_1\oplus\ldots\oplus\alpha_n=1}x_1^{\alpha_1}\ldots
 x_n^{\alpha_n}.\]
 Let $c_1,\ldots,c_n\in Z_k$ and $c_1\oplus\ldots\oplus c_n=1$. We can pick $t=f(x_{m+1}=c_{m+1},\ldots,x_n=c_n)$ and $r=c_{m+1}\oplus\ldots\oplus c_n$. Assume, without loss of generality, that $r=1$. Then we have 
 \[t= \bigoplus_{\alpha_1\oplus\ldots\oplus\alpha_m=0}x_1^{\alpha_1}\ldots
 x_m^{\alpha_m}.\]
 It must be shown that $M=Ess(t)$. By symmetry, it suffices to show that $x_1\in Ess(t)$. Let $c_2,\ldots,c_m\in Z_k$ be $m-1$ values, such that $c_2\oplus\ldots\oplus c_m=0$. Then we have 
 \[t(0,c_2,\ldots,c_{m})=1\quad\mbox{and}\quad t(1,c_2,\ldots,c_m)=0,\]
 which establishes that $x_1\in Ess(t)$.
  \end{case}
 \begin{case}[2]
$gap(f)=2$, $2<k< n$. 

 Theorem 2.1 from \cite{ros} implies that $Ess(f)=X_n$ and  $f$ is a symmetric function. According to  Theorem 4.1 \cite{sht_3}, it follows that all the non-empty subsets of $X_n$ are  separable in $f$.
 \end{case}
\begin{case}[3]  $gap(f)=2$, $n=3$ and $k\ge 3$.

 Using Theorem 5.1 and Proposition 5.1 in \cite{sht_2}, one can show that  $Ess(f)=X_n$ and  $M\subseteq X_n\Rightarrow M\in Sep(f)$, analogous to  
  Case 1.
\end{case}
\begin{case}[4] $gap(f)=2$, $4\le n\le k$. 

If  $f$ is a symmetric function then we are done because of  Theorem 4.1 in \cite{sht_3} and if $f$ is not a symmetric function then 
the proof is done as a part of   {\it Case 6}, given below.
\end{case} 
\begin{case}[5] $gap(f)=n$, $3\le n\le k$.

From Theorem 3.1 \cite{sht_2} it follows that $f$ is represented in the following form:
\begin{equation}\label{eq1}f=a_0\big[\bigoplus_{\alpha\in Eq_k^n}{\mathbf x}^\alpha\big]\oplus \big[\bigoplus_{\beta\in Dis_k^n} a_r{\mathbf x}^\beta\big],\end{equation} 
where 
 \[Eq_k^s=\{\gamma\in Z_k^s\ |\  \exists i,j,\  1\le i<j\le s,\ \gamma_i=\gamma_j\},\]
 $\gamma=\gamma_1\ldots\gamma_s$,\ ~ 
${\mathbf x}^\gamma=x_1^{\gamma_1}x_2^{\gamma_2}\ldots x_s^{\gamma_s}$
 and  
  $Dis_k^s=Z_k^s\setminus Eq_k^s$, for   $s$, $s\ge 2$. Moreover, there exist  at least two distinct coefficients  among $a_r\in Z_k$ for $r\in \{0,1,\ldots,k^n-1\}$.

It is easy to see that $Ess(f)=X_n$. Let $M$ be an arbitrary non-empty set of essential variables in $f$. We have to show that $M\in Sep(f)$. Without loss of generality let us assume that $M=\{x_1,\ldots,x_m\}$, $1\le m\le n$. If $m=n$ or $m=1$, we are clearly done. Let $1< m< n$ and let $\beta_{m+1},\ldots,\beta_n\in Z_k$ be $n-m$ constants such that the subfunction  $f_1=f(x_{m+1}=\beta_{m+1},\ldots x_n=\beta_n)$ is not a constant function.

By symmetry, it suffices to show that $x_1\in Ess(f_1)$. Without loss of generality let us assume that $a_0\neq b_1$, where $b_1=f_1(\gamma_1,\gamma_2,\ldots,\gamma_m)$ for some $\gamma_1,\ldots,\gamma_m\in Z_k$. Then, (\ref{eq1}) implies  
\[b_1=f_1(\gamma_1,\gamma_2,\ldots,\gamma_m)\neq a_0=f_1(\gamma_2,\gamma_2,\ldots,\gamma_m),\]
which shows that $x_1\in Ess(f_1)$ and  $Ess(f_1)=M$, and hence, $M\in Sep(f)$.
\end{case}
\begin{case}[6] $gap(f)=p$, $2\le p<n$, and $4\le n\le k$.

If $p>2$ then according to Theorem 3.4 in \cite{sht_2},  there exist two functions $h$ and $g$, such that $f=h \oplus g$, where  $g\in G_{n,k}^n$ and $ess(h)=n-p$. 
Moreover,  $g_{i\leftarrow j}=0$ for all $i$ and $j$,    $1\le j<i\le n$. The same representation of $f$ is established when $p=2$, in Theorem 4.2 \cite{sht_2}. 
Without loss of generality, let us assume that  
 $Ess(h)=\{x_{1},\ldots,x_{n-p}\}$.

 Clearly, $Ess(f)=X_n$ and according to (\ref{eq1}) we can pick $g=u\oplus v$, where 
\begin{equation}\label{eq3}u=\big[\bigoplus_{\beta\in Dis_k^n} a_r{\mathbf x}^\beta\big]\quad\mbox{and}\quad  v=a_0\big[\bigoplus_{\alpha\in Eq_k^n}{\mathbf x}^\alpha\big].
\end{equation}

Let 
  $x_i,x_j\in X_n$, $i> j$, be two arbitrary essential variables in $g$. For simplicity, say $i=n$ and $j=n-1$. Then we have 
\begin{equation}\label{eq4} g_{i\laa j}=u_{i\laa j}\oplus v_{i\laa j}=0\oplus a_0\big[\bigoplus_{\delta\in Z_k^{n-2}}{\mathbf {\hat x}}^\delta\big]=a_0.
\end{equation}
Since $g_{i\leftarrow j}=0$ for all $i$ and $j$,    $1\le j<i\le n$, it follows $v_{i\laa j}=a_0=0$. Consequently, $v=0$ and $g=u$.
Let $M$ be a non-empty set of essential variables in $f$. 
$M$ has to be separable set in $g$, according to {\it Case 5} and if $M\cap Ess(h)=\emptyset$ then it is also separable in $f$.

 We have to prove that $M\in Sep(f)$ in all other cases. We argue by induction on $n$ - the number of essential variables in $f$ and $g$.

Let $n=4$. This is our basis for induction.

First, let $|M|=2$ and $p=2$. 
Clearly, if $M\subseteq Ess(h)$ then (\ref{eq3}) and (\ref{eq4}) show that $M\in Sep(f)$. 
Next, let us assume that $M=\{x_1,x_3\}$ and $Ess(h)=\{x_1,x_2\}$. Let $c_2, c_4\in Z_k$ be two constants, such that $Ess(t_1)=\{x_1,x_3\}$, where $t_1=g(x_2=c_2,x_4=c_4)$. Clearly, $x_3\in Ess(f_1)$, where $f_1=f(x_2=c_2,x_4=c_4)$.
Let $h_1=h(x_2=c_2)$.
 If $x_1\notin Ess(h_1)$   then $x_1\in Ess(f_1)$ and obviously, $M\in Sep(f)$. If $x_1\in Ess(h_1)$ then $f_1= h_1(x_1)\oplus t_1(x_1,x_3).$ According to (\ref{eq3}) and (\ref{eq4}) there is a constant $c_3\in Z_k$, such that $Ess(t_1(x_3=c_3))=\emptyset$. Hence, $x_1\in Ess(f_1(x_3=c_3))$ and $M\in Sep(f)$, again.

Second, let $|M|=3$ and $p=2$, and $Ess(h)=\{x_1,x_2\}$.

Let $x_1\notin M$. Then there is a constant $c_1\in Z_k$, such that $x_2\in Ess(h_2)$, where $h_2=h(x_1=c_1)$. Thus, (\ref{eq3}) implies that $\{x_2,x_3,x_4\}= Ess(f_2)$, where $f_2=f(x_1=c_1)$ and $M\in Sep(f)$, again.

Let $x_4\notin M$. Then there is a constant $d_4\in Z_k$, such that $x_3\in Ess(t_2)$, where $t_2=g(x_4=d_4)$. Clearly, $x_3\in Ess(f_3)$, where $f_3=f(x_4=d_4)$. According to (\ref{eq3}) and (\ref{eq4}),  we have 
$f_3(x_3=d_4)= h(x_1,x_2)\oplus a_0$,
which shows that $\{x_1,x_2,x_3\}= Ess(f_3)$, and hence $M\in Sep(f)$.

One can similarly argue  if  $p=3$ and $n=4$.

Let us assume that for some natural number $l$, $l\ge 4$, if $n<l$, $2\le p$, $l< k$ and $f\in G_{p,k}^n$, then all non-empty sets of essential variables in $f$ are separable.

Let us pick $n=l$.
 According to Lemma \ref{l2}, there is a strongly essential variable $x_i$, $1\le i\le l$ in $g$,   and let $c_i\in Z_k$ be a constant, such that $X_l\setminus \{x_i\}= Ess(g(x_i=c_i))$. Without loss of generality, let us assume that $i=l$ and $c_i=k-1$. Using (\ref{eq3}), it is easy to see that 
 \begin{equation}\label{eq5}
 t_3=g(x_l=k-1)=\big[\bigoplus_{\beta\in Dis_{k-1}^{l-1}} b_r{\mathbf{\tilde x}}^\beta\big],
 \end{equation}
and the coefficients  $b_r$ linearly depend on $a_0,\ldots, a_{k^l-1}$, and ${\mathbf{\tilde x}}^\beta=x_1^{\beta_1}\ldots x_{l-1}^{\beta_{l-1}}$.
By $p\ge 2$ we may reorder the variables in $h$ such that $Ess(h)=\{x_1,\ldots,x_{l-p}\}$ with $l-p<l-1$.

 Then we can pick
$f_4=f(x_l=k-1)=h\oplus t_3$.
It must be shown that $Ess(f_4)=X_{l-1}$. Since $p\ge 2$, it follows that $N=Ess(f_4)\setminus Ess(h)\neq\emptyset$. Next, using (\ref{eq3}), one can show  that $N\in Sep(t_3)$ and $N\in Sep(f_4)$. According to (\ref{eq4}) we have 
\[N\cap Ess(f_4(x_i=k-1))=\emptyset,\]
 for all $i=1,\ldots, l-p$. Now, Lemma \ref{t1.6} implies $N\cup \{x_1,\ldots,x_{l-p}\}\in Sep(f_4)$. Hence, $Ess(f_4)=X_{l-1}$. According to (\ref{eq5}) it follows that $f_4\in G_{p,k-1}^{l-1}$. 

Therefore the inductive assumption may be applied to  $f_4$,  yielding  $M\in Sep(f_4)$, and hence,  $M\in Sep(f)$. 
\end{case} \qed
 \end{proof}

\section{Minor decision diagrams of functions}\label{sec3}

\begin{figure}[hbt]
\centering
\includegraphics[trim = 0 0 0 0]{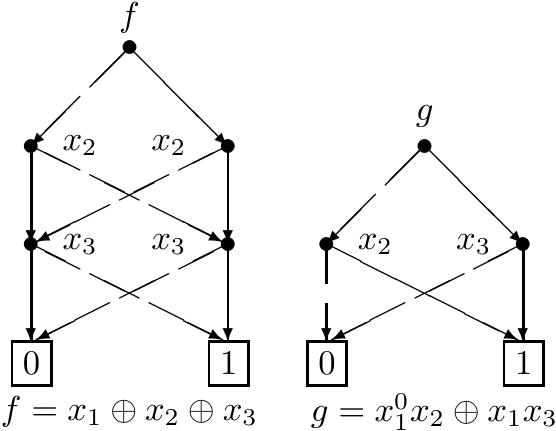}
\caption{BDD for $f$ and $g$ under the natural ordering of variables.}\label{f6}
\end{figure}

 Intuitively, it seems that a function $f$ has high complexity if all its sets of essential variables  are separable, because the  variables from separable sets remain essential after assigning constants   to other variables (see \cite{sh5}).
For example, when assigning Boolean constants to some variables of a Boolean function, then a natural complexity measure is the size of its Binary Decision Diagrams (BDDs), which also depend on the variable ordering (see \cite{and,bra}).
Each path from the root (function node) to a terminal node (leaf) of BDD is called an \emph{implementation} of $f$.
In \cite{sh5} we count the subfunction complexities  $imp(f)$, $sub(f)$ and $sep(f)$  of all implementations obtained under all $n!$ variable orderings, subfunctions, and separable sets  of   $n$-ary  Boolean functions for $n$, $n\le 5$.    
\begin{example}\label{ex5}
Let $f=x_1\oplus x_2\oplus x_3\ (mod\  2)$ and 
$g=x_1^0x_2\oplus x_1x_3\ (mod\ 2)$ be two Boolean functions. Figure \ref{f6} presents their BDDs.
All sets of essential variables in $f$ are separable, whereas the set $\{x_2,x_3\}$ is inseparable in $g$. Obviously, the BDD of the function $f$ is extremely complex with respect to the numbers of its subfunctions and implementations (see \cite{sh5}), whereas the  BDD of $g$ is very simple.
Thus we have $imp(f)=48$, $sub(f)=15$, $sep(f)=7$ and $imp(g)=28$, $sub(g)=11$, and $sep(g)=6$.
\end{example}
 
 This example shows that the functions having non-empty inseparable sets of essential variables must be  quite simple with respect to the subfunction complexities, whereas 
 Theorem \ref{t2} shows that the functions without inseparable sets of variables  have trivial arity gap, which means that they must be with high minor complexity.

 Roughly spoken, the  complexity of functions, is a mapping (evaluation) $Val:P_k^n\to\mathbb N$ with $Val(x)=c$ for all $x\in X$ and for some natural number $c\in \mathbb N$, called the \emph{initial value} of the complexity, and $Val(f)\ge c$ for all $f\in P_k^n$.

The  concept of complexity of functions is based on the "difficulties"   when  computing several resulting objects as subfunctions, implementations, separable sets, values, superpositions,  etc. 

As mentioned, we have used the computational complexities $sub(f)$, $imp(f)$ and $sep(f)$ in \cite{sh5} to classify the functions from the algebra  $P_k^n$. These complexities are invariants under the  action of the  groups $SB_k^n$, $IM_k^n$ and $SP_k^n$.

\begin{figure}
\centering
\includegraphics[trim = 20 0 0 0]{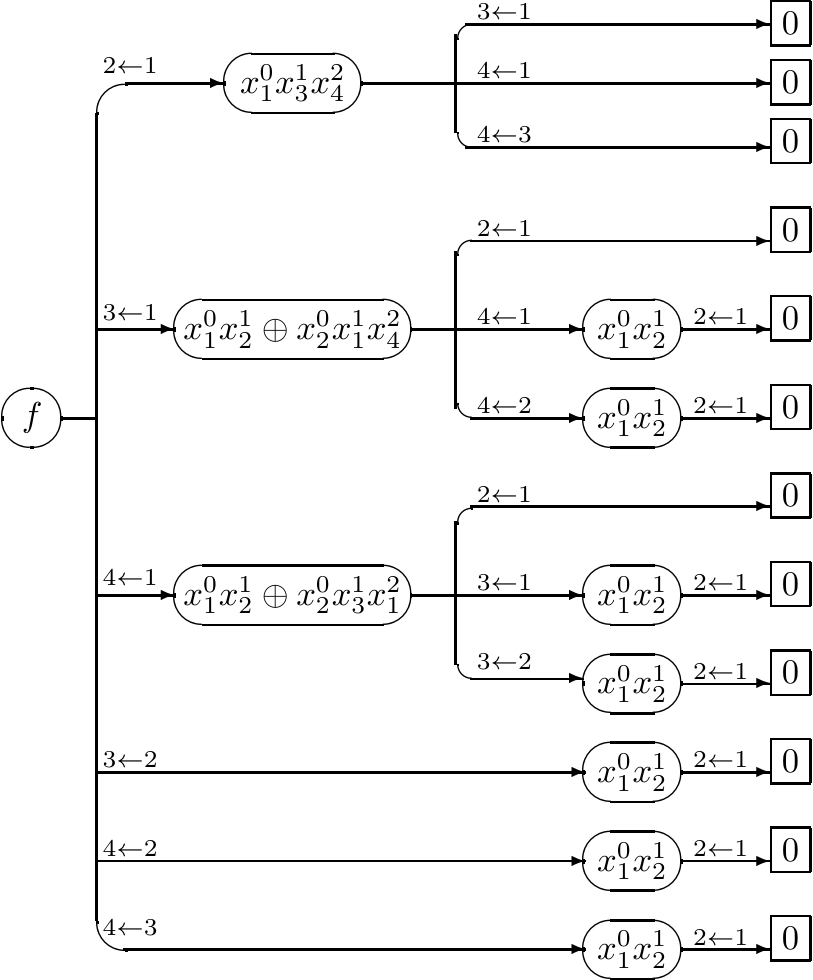}
\caption{Minor decomposition tree of $f=x_1^0x_2^1\oplus x_2^0x_3^1x_4^2\ (mod\ 3).$}\label{f1}
\end{figure}

 {Figure} \ref{f1}  shows the minor decomposition tree, constructed for the function 
$f=x_1^0x_2^1\oplus x_2^0x_3^1x_4^2\ (mod\ 3)$,  
 which essentially depends on all of its four variables $x_1, x_2, x_3$ and $x_4$.
The node at the left,  labelled  $f$ - is the \emph{function} node. 
 The nodes represented as ovals and labelled with  minor names  are the \emph{internal 
(non-terminal)} nodes, and the rectangular nodes (leaves of the tree) are the 
\emph{terminal} nodes. The terminal nodes are labelled  with the same name of a function (atomic minor) from $P_k^1$ (according to Theorem \ref{t3}).   
The terminal and non-terminal nodes in the  MDT for a function $f$, essentially depending on 
$n$ variables, are disposed   into maximum $n-1$ layers of the tree.  The $i$-th layer consists of   names of all the distinct minors of order $i$, for $i=1,\ldots,n-1$.

\begin{figure}[hbt]
\centering
\includegraphics[trim = 0 0 0 0]{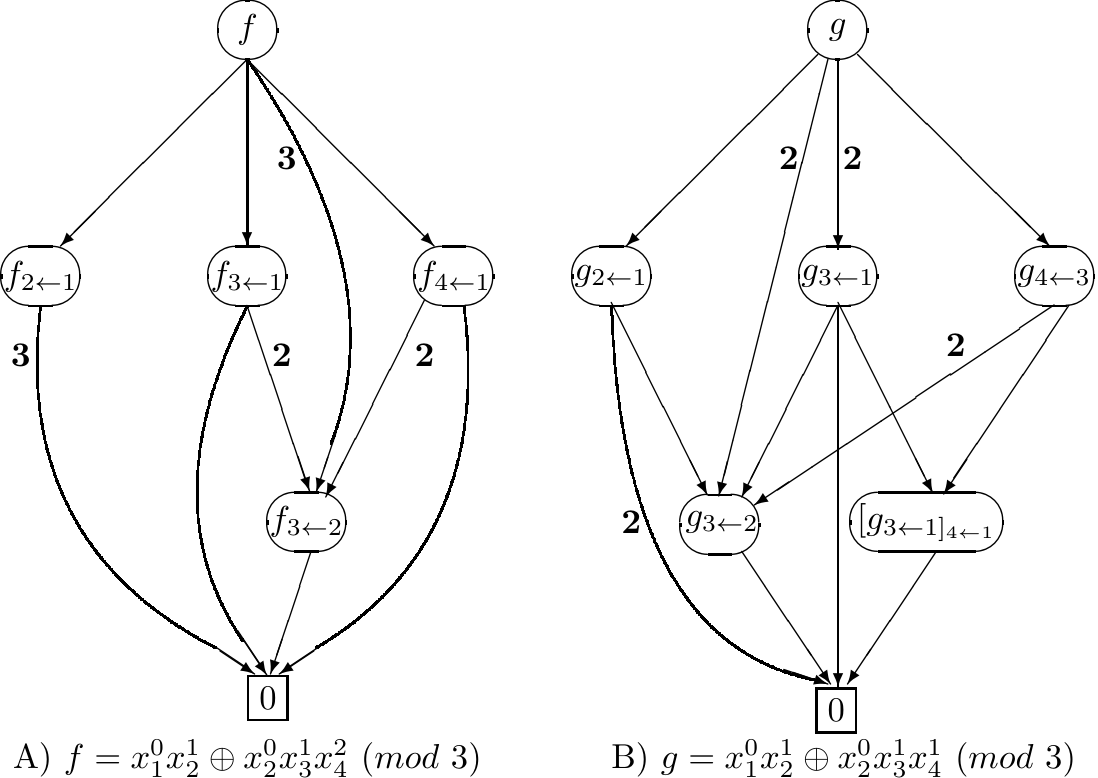}
\caption{Minor decision diagrams}\label{f2}
\end{figure}

We introduce  the \emph{minor decision diagrams} (MDDs) for $k$-valued
 functions constructed by reducing their \emph{minor 
decomposition trees} (MDTs). 
Let $f$ be a $k$-valued function.
The \emph{minor decision diagram (MDD)} of  $f$ is obtained from 
the corresponding MDT by \emph{reductions} of its nodes and edges applying of the following  
rules, starting from the MDT and continuing until neither rule can be applied:

{\bf Reduction rules}
\begin{enumerate}
\item[$\bullet$] If two edges have equivalent (as mappings) labels of their nodes they are merged. 
\item[$\bullet$] If two nodes  have equivalent labels, they are merged.
\end{enumerate}
Each edge $e=(v_1,v_2)$ in the diagram is supplied with a label $l(v_1,v_2)$, (written as bold in Figure \ref{f2} {A)}, which presents the number of the merged edges of the MDT, connecting the nodes $v_1$ and $v_2$ in MDT. If two nodes in MDT are connected with unique edge then this edge is presented in MDD without label, for brevity. For example, such pairs in  Figure \ref{f1} are $(f,f_{2\laa 1})$, $(f,f_{3\laa 1})$, $(f,f_{4\laa 1})$, $(f_{3\laa 1},0)$, $(f_{3\laa 2},0)$ and $(f_{4\laa 1},0)$.

So, the MDD of $f$ is an acyclic directed graph, with unique function node and according to Theorem \ref{t3}, with unique terminal node. Clearly, the MDD and MDT are uniquely determined by the function $f$.

\begin{example}\label{ex12}
Let us build the MDDs of the following two functions
$f=x_1^0x_2^1\oplus x_2^0x_3^1x_4^2\ (mod\ 3)$  and $g=x_1^0x_2^1\oplus x_2^0x_3^1x_4^1\ (mod\ 3)$, using the reduction rules.

 {Figure} \ref{f2} {A)} shows the MDD of the function $f$,
 obtained from its MDT, given in Figure \ref{f1}, after applying the reduction rules.
 Figure  \ref{f2} {B)} presents the MDD of $g$.
 The identification minors of $f$ and $g$ are:\\
\begin{tabular}{ll}
 $f_{2\laa   1}= x_1^0x_3^1x_4^2\ (mod\ 3)$,& 
$f_{3\laa    1}= x_1^0x_2^1\oplus x_2^0x_1^1x_4^2\ (mod\ 3)$,\\ 
$f_{4\laa    1}= x_1^0x_2^1\oplus x_2^0x_3^1x_1^2\ (mod\ 3)$,& $f_{3\laa    2}= x_1^0x_2^1\ (mod\ 3)$,\\
\\[.1ex] 
 $g_{2\laa   1}= x_1^0x_3^1x_4^1\ (mod\ 3)$,& 
$g_{3\laa    1}=x_1^0x_2^1\oplus x_2^0x_1^1x_4^1\ (mod\ 3)$, \\
 $g_{4\laa    1}=x_1^0x_2^1\oplus x_2^0x_3^1x_1^1\ (mod\ 3)$, &
   $g_{4\laa    3}=x_1^0x_2^1\oplus x_2^0x_3^1\ (mod\ 3)$, \\
$[g_{2\laa   1}]_{4\laa    3}= x_1^0x_3^1\ (mod\ 3)$, &
$[g_{3 \laa   1}]_{4\laa    1}=x_1^0x_2^1\oplus x_2^0x_1^1\ (mod\ 3)$, \\
$g_{3\laa    2}=x_1^0x_2^1\ (mod\ 3)$. &
\end{tabular}\\
Clearly,
$f_{3\laa 2}=f_{4\laa 2}=f_{4\laa 3}=[f_{{4\laa 1}]_{3\laa 1}}=[f_{{4\laa 1}]_{3\laa 2}}=[f_{{3\laa 1}]_{4\laa 1}}=[f_{{3\laa 1}]_{4\laa 2}}=x_1^0x_2^1\ (mod\ 3)$.
The  minors $f_{i\laa 1}$ for $i=2,3,4$ are of order $1$, and the last minor $f_{3\laa 2}$ is of order $2$.

The label of the edge  $(f,f_{3\laa 2})$ is {\bf 3} because there are three identification minors, namely $f_{3\laa 2}$, $f_{4\laa 2}$ and $f_{4\laa 3}$  of $f$ which  are equivalent to $f_{3\laa 2}$ (see the last three branches of the MDT in Figure \ref{f1}). In a similar way we count the labels of the  edges in Figure \ref{f2} {B)}.
Note that $g_{3\laa 1}\equiv g_{4\laa 1}$ and $[g_{{2\laa 1}]_{4\laa 3}}\equiv g_{3\laa 2}$, which implies that the nodes labelled with these minors must be merged  in the MDD of $g$. 
The functions $f$ and $g$ are very close (the difference  is that  $x_4^2$ in $f$ is changed to $x_4^1$ in $g$)  in their formula representations, but the diagram of $g$ is more complex as we see  in Figure \ref{f2} {B)}.
 \end{example}

The size of the MDD and the minor complexities are determined   by the function, being represented. The "scalability" of the diagram is an important measure of the computational complexity of the function. We are going to formalize this problem and establish a method for classification of   functions by the minor complexities.

First,  the number  $mnr(f)$ of all the minors of a function $f$ is a complexity measure, which can be used to evaluate the MDD of $f$. Namely, it counts the size (number of terminal and non-terminal nodes) of the MDD. M. Couceiro, E.  Lehtonen and T. Waldhauser have studied similar evaluation, named "parametrized arity gap"   in \cite{er11,er1}, which characterizes the sequential identification minors of a function.

 We  define two new complexity measures which  count the number of minors and the number of ways to obtain these minors. Our goal is to  classify functions in  finite algebras by these complexities. 
\begin{definition} \label{13} 
Let $f\in P_k^n$ be a $k$-valued function. Its \emph{$cmr$-complexity} $cmr(f)$ is defined as follows:
\begin{enumerate}
\item[(i)]$cmr(f)=1$ if $ess(f)\le 1$;
\item[(ii)]$cmr(f)=2$ if $ess(f)=2$;
\item[(iii)]$cmr(f)=\sum_{j<i,\ x_i,x_j\in Ess(f)}cmr(f_{i\laa j})$ if $ess(f)\ge 3$.
\end{enumerate}
\end{definition}
 
 The minors $f_{i\laa j}$ with $i<j$ are excluded  because   $f_{i\laa j}\equiv f_{j\laa i}$. 
  The minor complexity $cmr$ can be  inductively calculated  using the MDDs of the functions as it is shown in Algorithm \ref{alg1}, below. 
  We start to assign $cmr$-complexity equals to $1$ for the terminal node, according to (i) of Definition \ref{13}.  
 Next, we calculate the   $cmr$-complexity of the minors of $f$ with lower order,  applying (ii) and (iii)  of Definition \ref{13}.

\begin{example}\label{ex14}
We now count the $cmr$-complexity of the functions $f$ and $g$ from Example \ref{ex12}, using their MDDs given in Figure \ref{f2} {A)} and {B)}, respectively. There is one minor ($f_{3\laa 2}$)  of  order $2$ and three minors of order $1$. Thus we have 
$cmr(f_{3\laa 2})=2$, $cmr(f_{2\laa 1})=1.3=3$, $cmr(f_{3\laa 1})=1.1+2.2=5$, and $cmr(f_{4\laa 1})=1.1+2.2=5$. Again, using (ii) and (iii) of Definition \ref{13} we obtain 
$cmr(f)=3+5+5+3.2=19$.

The MDD of $g$ in Figure \ref{f2} {B)} shows that: \\
\begin{tabular}{ll}
$cmr(g_{3\laa 2})=cmr([g_{{3\laa 1}]_{4\laa 1}})=2$,& $cmr(g_{2\laa 1})=2.1+1.2=4$,\\
 $cmr(g_{3\laa 1})=1.2+1.1+1.2=5$,& $cmr(g_{4\laa 3})=2.2+1.2=6$,\\ and $cmr(g)=1.4+2.2+2.5+1.6=24$. &
 \end{tabular}\\ 
We clearly have: $mnr(f)=5$ and $mnr(g)=6$.

 It is clear that the set $M=\{x_1,x_3,x_4\}$ is inseparable in both $f$ and $g$.

  \end{example}

\begin{theorem}\label{t14}
Let $f\in P_k^n$ with $2\le ess(f)=n\le k$. Then
\begin{enumerate}
\item[(i)]$\frac{n(n-1)}{2}\le cmr(f)\le \frac{n!(n-1)!}{2^{n-2}}$;
\item[]
\item[(ii)]$1\le mnr(f)\le \frac{n!(n-1)!}{2^{n-2}}$.
\end{enumerate}
\end{theorem}
\begin{proof}
The maximum number of  minors of order $1$ for a function $f$ is equal to $\binom{n}{2}$. Using this   as an inductive basis one can show that the maximum number of minors of order $m$, $1\le m\le n-1$ is equal to $\binom{n}{2}\binom{n-1}{2}\ldots \binom{n-m+1}{2}$. Pick  
\[f=x_1(x_2\oplus 1)\ldots (x_n\oplus (n-1))\ (mod\ k)\]
 with $n\le k$. One can inductively prove that  all the  minors of $f$ are pairwise distinct, which shows that $f$ reaches  the upper bound of (i) and (ii), and establishes the right inequalities in  (i) and (ii).

The lower bound in (ii) is clear. If $n\le k$ then the minimum number of minors in a function depending essentially on $n$ variables is reached to the functions $f$ from $G_{n,k}^n$ with $gap(f)=n$. It follows that $f$ is represented as in (\ref{eq1}), which shows that $cmr(f)=\binom{n}{2}=\frac{n(n-1)}{2}$. 
\end{proof}

\begin{remark}\label{r14}
Note that the upper bound of $cmr$ and $mnr$ is independent on  $k$   and hence, it is satisfied  in all the possible cases  for $2\le n\le k$.

Later,  we shall discuss lower and upper bounds of $cmr$ and $mnr$ when $k<n$, and   for Boolean functions. A  hypothesis, here is that the upper bound for (ii) is unreachable for   $k$ and $n$ if $k<n$. 
\end{remark}

\section{Equivalence  relations with respect to minor complexities}\label{sec31}

Many of the problems in the applications of the $k$-valued logic are compounded because of the large number of the functions, namely $k^{k^n}$. Techniques which involve enumeration of functions can only be used if $k$ and $n$ are trivially small. A common way for extending the scope of such enumerative methods is to classify the functions into equivalence classes by some natural equivalence relation.

Let $S_A$ denote the symmetric group of all permutations of the non-empty set $A$, and let   $S_m$ denote   the   group $S_{\{1,\ldots,m\}}$ for a natural number $m$, $m\ge 1$.

A \emph{transformation} $\psi:P_k^n\longrightarrow P_k^n$ is  an
$n$-tuple of $k$-valued functions
$\psi=(g_1,\ldots,g_n)$, $g_i\in P_k^n$, $i=1,\ldots,n$
acting on any function $f=f(x_1,\ldots,x_n)\in P_k^n$ as follows
$\psi(f)=f(g_1,\ldots,g_n)$.
Then the  composition of two transformations $\psi$ and $\phi=(h_1,\ldots,h_n)$
is defined as follows
\[\psi\phi=(h_1(g_1,\ldots,g_n),\ldots,h_n(g_1,\ldots,g_n)).\]
The set of all transformations of  $P_k^n$ is the \emph{universal monoid $\Omega_k^n$} with unity - the identical transformation $\epsilon=(x_1,\ldots,x_n)$. When taking only invertible transformations we obtain the \emph{universal group}  $C_k^n$ isomorphic to the symmetric group $S_{Z_k^n}$.
 The groups consisting of invertible transformations of $P_k^n$ are called \emph{transformation groups} (sometimes termed \emph{permutation groups}). 

Let $\simeq$, $\simeq\subseteq P_k^n\times P_k^n$ be an equivalence relation on the algebra $P_k^n$.
 Since $P_k^n$ is a finite algebra of $k$-valued functions, the equivalence relation $\simeq$ makes a partition of the algebra in a finite number equivalence classes $P_1,\ldots,P_r$.

A mapping $\varphi:P_k^n\longrightarrow P_k^n$ is called a \emph{transformation preserving $\simeq$} if $f\simeq \varphi(f)$ for all $f\in P_k^n$.
Taking only invertible transformations which preserve $\simeq$, we get the group $G_{\simeq}$ of all transformations preserving $\simeq$. The \emph{ orbits} (also called \emph{$G_{\simeq}$-types}) of this group  are exactly the classes $P_1,\ldots,P_r$. 
The number $r$ of orbits of a group $G_{\simeq}$ of transformations  is denoted   $t(G_{\simeq})$.
Since we want to classify functions from $P_k^n$ into equivalence classes by $\simeq$, three natural problems occur, namely:
\begin{enumerate}
\item[](i) Calculate the number $t(G_{\simeq})$ of $G_{\simeq}$-types; 
\item[](ii) Count the number of functions in different equivalence classes, i.e. compute the cardinalities of the sets  $P_1,\ldots,P_r$; 
\item[](iii) Make a catalogue (list) of functions belonging to different $G_{\simeq}$-types.
\end{enumerate}

Let $f\in P_k^n$ and let   $nof(f)$ denote the normal form obtained by applying the reduction $\rhd$ on $f$. According to Theorem \ref{t3}, the normal form $nof(f)$ is unique and $nof(f)\in P_k^1$. Thus, our first natural equivalence is defined as follows:
\begin{definition}\label{d18} Let $f$ and $g$ be two functions  from $P_k^n$.
We say that $f$ and $g$ are 
 {$nof$-equivalent} (written $f\simeq_{nof} g$)  if 
$nof(f)=nof(g)$.
\end{definition}
The transformation group induced by $nof$-equivalence is denoted   $NF_k^n$. The transformations in $NF_k^n$ preserves $\simeq_{nof}$, i.e.   $nof(g)=nof(\psi(g))$ for all $g\in P_k^n$ and $\psi\in NF_k^n$. Since the atomic minors (labels of terminal nodes in MDD)  depend on at most one essential variable, it follows  that $t(NF_k^n)=|P_k^1|=k^k$.  These transformations  involve permuting variables, only (see Theorem \ref{t17}, below). The $nof$-equivalence is independent on the $cmr$-complexity of functions, defined by reduction via minors. For example, the functions $f=0$ and $g=x_1^0x_2\oplus  x_1x_2x_3^0(mod\ 2)$ are $nof$-equivalent, but $cmr(f)=1$ and $cmr(g)=6$.

By analogy with the ordered decision diagrams  \cite{bra,sh5}, we   define several  equivalence relations in $P_k^n$, which allow us to classify the functions by the complexity of their MDDs.
\begin{definition}\label{d2.4} Let $f$ and $g$ be two functions  from $P_k^n$.
We say that $f$ and $g$ are 
 {$cmr$-equivalent} (written $f\simeq_{cmr} g$)  iff:
\begin{enumerate}
\item[(i)] $ess(f)\le 1$ $\implies$  $ess(f)=ess(g)$;
\item[(ii)] 
 $ess(f)\ge 2$ $\implies$ $ess(f)=ess(g)$ and there exists a permutation $\sigma$ of the set $\{1,\ldots,n\}$ such that 
$f_{i\laa j}\simeq_{cmr} g_{\sigma(i)\laa \sigma(j)}$
for all $j,i$, with $x_i,x_j\in Ess(f)$, $j<i$. 
\end{enumerate} 
\end{definition}
Let  $CM_k^n$ denote the transformation group  preserving the equivalence 
$\simeq_{cmr}$.
Note that if $ess(f)\le 1$ then $Mnr(f)=\emptyset$. Hence, if $ess(f)=ess(g)\le 1$ then $f\simeq_{mnr} g$.
$MN_k^n$ denotes the transformation group which preserves the equivalence $\simeq_{mnr}$.

Next we define another equivalence based on the number of minors (size of  MDD) in a function.
 \begin{definition}\label{d17} Let $f$ and $g$ be two functions  from $P_k^n$.
We say that $f$ and $g$ are \emph{$mnr$-equivalent} (written $f\simeq_{mnr} g$)  if  
$mnr_m(f)=mnr_m(g)$
for all $m$,  $0\le m\le ess(f)-1$. 
\end{definition}
 Note that $f\simeq_{mnr} g$ or $f\simeq_{nof} g$ do not imply $ess(f)=ess(g)$, which can be seen by the following functions: $f=x_1^0x_2^1 (mod\ 3)$ and $g=x_1^0x_2^1x_3^2 (mod\ 3)$. Clearly,   $f\simeq_{mnr} g$ and $f\simeq_{nof} g$, but  $ess(f)=2$, and $ess(g)=3$.
  \begin{theorem}\label{t166}~
  \begin{enumerate}
\item[(i)]$f\simeq_{cmr} g\ \implies cmr(f)=cmr(g)$;
\item[(ii)] $f\simeq_{cmr} g\ \implies f\simeq_{mnr} g$.
\end{enumerate}
 \end{theorem} 
 \begin{proof}
 We argue by induction on the number   $n=ess(f)$.
 
 If $ess(f)\le 2$ (basis for induction) then we are clearly done. Assume that (i) and (ii) are satisfied when $n<s$ for some natural number $s$, $s>2$.  Let $n=s$ and $f\simeq_{cmr} g$. Then our inductive assumption implies \[cmr(f)=\sum_{j<i,\ x_i,x_j\in Ess(f)}cmr(f_{i\laa j})=\]
 \[ \sum_{u<v,\ x_u,x_v\in Ess(g)}cmr(g_{u\laa v})=cmr(g),\]  and $mnr_m(f_{i\laa j})=mnr_m(g_{u\laa v})$, where $u=\pi(i)$ and $v=\pi(j)$ for some $\pi\in S_n$ and $m=0,\ldots, n-1$.~~~~~~
\end{proof}  
Thus, the complexity $cmr(f)$ is an invariant of the group $CM_k^n$, so that if $f\simeq_{cmr} g$ then $cmr(f)=cmr(g)$, and the complexity $mnr(f)$ is an invariant of the group $MN_k^n$, so that if $f\simeq_{mnr} g$ then $mnr(f)=mnr(g)$.
 
It is naturally to ask which   groups among "traditional" transformation groups are subgroups of the groups $NF_k^n$ or $CM_k^n$ and  which of these groups include $NF_k^n$, $MN_k^n$ or $CM_k^n$ as their subgroups. 

Let $\sigma:Z_k\longrightarrow Z_k$ be a mapping and $\psi_\sigma:P_k^n\longrightarrow P_k^n$ be a transformation of $P_k^n$ generated by $\sigma$ as follows 
$\psi_\sigma(f)(\o a)=\sigma(f(\o a))$
for all $\o a\in Z_k^n$.
 
\begin{theorem}\label{t16}
The transformation $\psi_\sigma$ preserves $\simeq_{cmr}$  if and only if $\sigma$ is a permutation of $Z_k$, $k>2$.
\end{theorem}
\begin{proof} 
First, let $\sigma\in S_{Z_k}$ be a permutation of $Z_k$. Let $f\in P_k^n$ be an arbitrary function. If $ess(f)\le 1$ then $ess(\psi_\sigma(f))=ess(f)$ and we are clearly done. Let $ess(f)=ess(g)=n\ge 2$ and let $i$ and $j$, $1\le j<i\le n$ be two arbitrary natural numbers. Then we have 
\[[\psi_\sigma(f)]_{i\laa j}(x_1,\ldots,x_n)=\sigma(f_{i\laa j}(x_1,\ldots,x_n)).\]
Since $\sigma$ is a permutation, it follows that $f_{i\laa j}\simeq_{cmr}[\psi_\sigma(f)]_{i\laa j}$ which shows that $f\simeq_{cmr}\psi_\sigma(f)$.

Second, let $\sigma$ be not a permutation of $Z_k$. Hence, there exist two constants $a_1$ and $a_2$ from $Z_k$ such that $a_1\neq a_2$ and $\sigma(a_1)=\sigma(a_2)$. Let $\o b=(b_1,\ldots,b_n)\in Z_k^n$, $n\ge 2$ be a vector of constants from $Z_k$. Then we define the following function from $P_k^n$:
\[
f(x_1,\ldots,x_n)=\left\{\begin{array}{ccc}
            a_1 \  &\  if \  &\  x_i=b_i\ for\ i=1,\dots,n \\
            a_2 &   & otherwise.
           \end{array}
           \right.
\]
Clearly, $Ess(f)=X_n$ and the range of $f$ consists of two numbers, i.e. $A=\{a_1,a_2\}$. Then $\sigma(A)=\{\sigma(a_1)\}$,  implies that 
$\psi_\sigma(f)(c_1,\ldots,c_n)=\sigma(a_1)$
for all $(c_1,\ldots,c_n)\in Z_k^n$. Hence, $Ess(\psi_\sigma(f))=\emptyset$, which shows that  $f\not\simeq_{cmr} \psi_\sigma(f)$ and $\psi_\sigma\notin CM_k^n$.~~~~
\end{proof}
Let $\pi\in S_n$ and $\phi_{\pi}:P_k^n\longrightarrow P_k^n$ be a transformation of $P_k^n$ defined as follows $\phi_{\pi}(f)(a_1,\ldots,a_n)=f(a_{\pi(1)},\ldots,a_{\pi(n)})$
for all $(a_1,\ldots,a_n)\in Z_k^n$.
\begin{theorem}\label{t17}
The transformation $\phi_{\pi}$ preserves the equivalence relations $\simeq_{cmr}$,  $\simeq_{mnr}$ and $\simeq_{nof}$ for all $\pi\in S_n$.
\end{theorem}
\begin{proof}
It suffices to show that  $\phi_{\pi}$ preserves $\simeq_{cmr}$ and $\simeq_{nof}$. 
 
Let $f\in P_k^n$ be a function and let us assume $Ess(f)=X_n$, $n\ge 2$. It must be shown  that $f\simeq_{cmr} g$ and $f\simeq_{nof} g$, where 
$g(a_1,\ldots,a_n)=f(\pi(a_1),\ldots,\pi(a_n))$
for all $(a_1,\ldots,a_n)\in Z_k^n$. Since $\pi$ is a permutation, we have 
\[f(a_1,\ldots,a_n)=g(\pi^{-1}(a_1),\ldots,\pi^{-1}(a_n)),\]
for all $(a_1,\ldots,a_n)\in Z_k^n$ and one can easily show that $f\simeq_{cmr} g$ and hence, $f\simeq_{cmr} \phi_{\pi}(f)$. Since $nof(f)=f(x_i,\ldots,x_i)$ and  $nof(f)=f(x_{\pi(i)},\ldots,x_{\pi(i)})$, it follows $nof(g)\equiv nof(g)$ and $f\simeq_{nof} g$.
\end{proof}
  
We  deal with "natural" equivalence relations which involve  variables of
functions. Such relations induce permutations of the domain $Z_k^n$ of the
functions. These mappings form a transformation group whose number of
equivalence classes  can be determined.
The restricted affine group (RAG) is defined as a subgroup of the symmetric
group on the direct sum of the module $Z_k^n$ of arguments of functions
and the ring $Z_k$ of their outputs. The group RAG permutes the direct
sum  $Z_k^n+Z_k$ under restrictions which preserve single-valuedness of all
functions from $P_k^n$ \cite{har2,lech1,str3}.

In the model of RAG an affine transformation $\psi$ operates on the domain or
space of inputs $\mathbf{x}=(x_1,\ldots,x_n)$ to produce the output
$\mathbf{y}=\mathbf{xA}\oplus \mathbf{c}$, which might be used as an input in
the function $f$. Its output $f(\mathbf{y})$ together with the function
variables $x_1,\ldots,x_n$ are linearly combined by a range transformation which
defines the image $g=\psi(f)$ of $f$ as follows:
\begin{eqnarray}\label{eq51} g(\mathbf{x})=\psi(f)(\mathbf{x})=f(\mathbf{y})\oplus a_1x_1\oplus\ldots\oplus
a_nx_n\oplus d= \nonumber\\
f(\mathbf{xA}\oplus \mathbf{c})\oplus \mathbf{a^tx}\oplus d,
\end{eqnarray}
where $d$ and $a_i$ for $i=1,\ldots,n$ are constants from $Z_k$.
Such a  transformation belongs to RAG if $\mathbf{A}$ is a non-singular matrix.

We want to extract basic facts for several subgroups of RAG which are
"neighbourhoods" or "relatives" of our transformation groups $NF_k^n$, $CM_k^n$ and $MN_k^n$. It is also interesting to compare these groups with the groups  $IM_k^n$,
$SB_k^n$ and $SP_k^n$, studied in \cite{sh5}.

First, a classification occurs when permuting arguments of functions. If $\pi\in S_n$   then
$\pi$ acts on variables by:
$\pi(x_1,\ldots,x_n)=(x_{\pi(1)},\ldots,x_{\pi(n)}).$
Each permutation generates a map on the domain $Z_k^n$. 

For example, the
permutation $\pi=(1,3,2)$ generates a permutation of the domain  $\{0,1,2\}^3$ of the functions from $P_3^3$. Then we have $\pi: 001\ra 010 \ra 100$ and in cyclic decimal
notation this permutation can be written as  
$(1,3,9)$. The remaining elements of $Z_3^3$ are mapped  according to the following cycles of $\pi$ in decimal notation - $(2,6,18)(4,12,10)$ $(5,15,19)(7,21,11)(8,24,20)$ $(14,16,22)(17,25,23)$. Note that each permutation from $S_n$ keeps  fixed all $k$ constant tuples from $Z_k^n$. In case of $Z_3^3$, these tuples $(0,0,0)$, $(1,1,1)$ and $(2,2,2)$ are presented by the decimal numbers $0,\ 13$ and $26$. 

$S_k^n$ denotes the transformation group induced by permuting of variables. 

Boolean functions of two variables are classified into twelve  $S_2^2$-classes \cite{har2}, as it is shown in Table \ref{tb11}.
\begin{table}[hbt]
\centering
\caption{The twelve classes in  $P_2^2$ under the permutation of arguments.} \label{tb11}
\begin{tabular}{|llll|}\hline\hline
~[$0$],& [$x_1^0x_2^0$], & [$x_1^0x_2,\ x_1x_2^0$],   & ~[$x_1,\ x_2$],\\ ~ [$x_1\oplus x_2$],&[$x_1\oplus x_2^0$],&  ~[$x_1^0\oplus x_1x_2,\ x_2^0\oplus x_1x_2$], & [$x_1\oplus x_1^0x_2$], \\ ~ [$x_1^0\oplus x_1x_2^0$],&
~[$x_1x_2$],& [$x_1^0,\ x_2^0$],&[$1$].\\
\hline\hline
\end{tabular} 
\end{table}

 M. Harrison  has determined  the cycle index of the group $S_2^n$  and using Polya's counting theorem he has counted the number of equivalence classes under permuting arguments (see \cite{har2} and Table \ref{tb1}, below).

The following proposition is obvious.
\begin{proposition}
 The transformation group  $S_k^n$ is induced by the equivalence relation $\equiv$ (see Definition \ref{d1}).
\end{proposition}
The subgroups of RAG, defined according to (\ref{eq51}) are
determined  by equivalence relations as it is shown in Table \ref{tb2}, where
$\mb{P}$ denotes a permutation matrix, $\mb{I}$ is the identity matrix,
$\mb{b\mbox{ and }c}$ are $n$-dimensional vectors over $Z_k^n$ and $d\in Z_k$.

\begin{table}[hbt]
\centering
\caption{The subgroups of RAG}\label{tb2}
\begin{tabular}{|l|l|l|}\hline\hline
Groups& Equivalence & Determination\\
&relations&\\ \hline
RAG & Affine transformation& $\mb{A}$-non-singular\\[.5ex]
 $GE_k^n$ & Genus & $\mb{A}=\mb{P}$, $\mb{a}=\mb{0}$\\[.5ex]
 $CF_k^n$ & Complement outputs& $\mb{A}=\mb{I}$, $\mb{a}=\mb{0}$, $\mb{c}=\mb{0}$
\\[.5ex]
    $G_k^n$  & Symmetry  types& $\mb{A}=\mb{P}$, $\mb{a}=\mb{0}$, $d=0$\\[.5ex]
    $LF_k^n$ & Add linear  function& $\mb{A=I}$, $\mb{c=0}$,  $d=0$\\[.5ex]
$CA_k^n$ & Complement  arguments& $\mb{A}=\mb{I}$, $\mb{a}=\mb{0}$, $d=0$\\[.5ex]
 $LG_k^n$ & Linear  transformation& $\mb{c}=\mb{0}$, $\mb{a}=\mb{0}$, $d=0$\\[.5ex]
   $S_k^n$ & Permute variables& $\mb{A}=\mb{P}$, $\mb{c}=\mb{0}$, $\mb{a}=\mb{0}$, $d=0$\\[.5ex]
 \hline\hline
\end{tabular}
\end{table}

It is naturally to ask which subgroups of RAG are subgroups of the group
$NF_k^n$, $CM_k^n$ and $MN_k^n$. Theorem \ref{t16} and Theorem \ref{t17} show that $CF_k^n$ and $S_k^n$ are subgroups of $CM_k^n$.  Theorem \ref{t166} shows that they must also be subgroups of $MN_k^n$. Clearly, $S_k^n\le NF_k^n$.
\begin{example}\label{ex19}
Let
$f=x_1\oplus x_2\oplus x_3(mod\ 3)$ {and} $g=x_1x_2\oplus x_1x_3\oplus x_2x_3(mod\ 3)$. 
Then
we have $f_{i\laa j}=2x_j\oplus x_m(mod\ 3)$ and $g_{i\laa j}=2x_jx_m\oplus x_jx_j(mod\ 3)$ where $\{i,j,m\}=\{1,2,3\}$. Clearly,
$f_{{i\laa j}_{j\laa m}}=g_{{i\laa j}_{j\laa m}}=0$, 
and hence  $f\simeq_{cmr} g$ and $f\simeq_{nof} g$.
One can show  that there 
is  no  transformation $\psi\in RAG$, defined as in (\ref{eq51}),  for which $g=\psi(f)$.
Consequently,  $CM_k^n\not\le RAG$, $NF_k^n\not\le RAG$ and $MN_k^n\not\le RAG$.
\end{example}
\begin{example}\label{ex20}
Let
$f=x_1^0x_2^1\oplus x_2^0x_3^1x_4^2(mod\ 3)\ \mbox{and}\  g=x_1^0x_2^1\oplus x_2^0x_3^1x_4^1(mod\ 3)$ be
the  functions from Example \ref{ex14} whose MDDs are given in Figure \ref{f2}. Let 
\begin{displaymath}
\mb{A}=\left(
\begin{array}{cccc}
1&0&0&0\\
0&1&0&0\\
0&0&1&0\\
0&0&0&2
\end{array}\right)
\end{displaymath}
 Then clearly, $f(\mb{x})=g(\mb{A}\mb{x})$ and hence, $f$ and $g$ belong to the same equivalence class under the transformation groups $LG_3^4$. 
 Let $\mb{c}=(0,0,0,1)$. Then we have $f(\mb{x})=g(\mb{x}.\mb{I}\oplus \mb{c})$, which shows that $f$ and $g$ belong to  the same equivalence class under the transformation group $CA_3^4$. One can   show that  $f\simeq_{nof} g$.
 Example \ref{ex14} shows that $f\not\simeq_{mnr} g$.
Consequently, $NF_k^n\not\le MN_k^n$, $LG_k^n\not\le MN_k^n$ and $CA_k^n\not\le MN_k^n$. Theorem \ref{t166} shows that  $NF_k^n\not\le CM_k^n$, $LG_k^n\not\le CM_k^n$ and $CA_k^n\not\le CM_k^n$.
\end{example}
\begin{example}\label{ex21}
Let
$f=x_1^0x_2\oplus x_1^1x_3\oplus x_1^2x_2^1x_3^0(mod\ 3)$ and $g=x_1^0x_2\oplus x_1^1x_3(mod\ 3)$ be
two  functions.
It is easy to see that $f_{i\laa j}=g_{i\laa j}$ for all $i,j$ with $1\le j<i\le 3$. Hence, $f\simeq_{cmr} g$ and $f\simeq_{nof} g$. Now, it is clear that each set of essential variables in $f$ is separable in $f$, but $\{x_2,x_3\}\notin Sep(g)$ which shows that $f\not\simeq_{sep} g$, i.e. $CM_k^n\not\le SP_k^n$ and  $NF_k^n\not\le SP_k^n$.
\end{example}
So, the next theorem summarizes     results  which determine the positions of the groups  $NF_k^n$, $CM_k^n$ and $MN_k^n$, with respect to the subgroups of RAG and the groups induced by subfunction complexities \cite{sh5}.
\begin{theorem}\label{t22} ~~

\begin{tabular}{lll}
(i) $CF_k^n\le CM_k^n$; &(ii) $S_k^n\le CM_k^n$;& (iii) $S_k^n\le NF_k^n$;\\
  (iv)   $NF_k^n\not\le RAG$;&(v) $CM_k^n\not\le RAG$; &
(vi) $LG_k^n\not\leqslant MN_k^n$;\\

 (vii) $CA_k^n\not\leqslant MN_k^n$;& (viii) $CA_k^n\not\leqslant NF_k^n$;&(ix) $LG_k^n\not\leqslant NF_k^n$;\\
  (x) $CM_k^n\not\le SP_k^n$;&
(xi) $CM_k^n\not\le SP_k^n$; & (xii) $CM_k^n\not\le NF_k^n$\\
(xiii)  $NF_k^n\not\le MN_k^n$.&&  
\end{tabular}
\end{theorem}
\begin{proof}
 (i) follows from Theorem \ref{t16}, (ii) and (iii)- from Theorem \ref{t17}, (iv) and (v) - from Example \ref{ex19}, (vi), (vii), (viii), (ix) and (xiii) - from Example \ref{ex20},   (x) and (xi)- from Example \ref{ex21}.

 To show (xii) let us pick $f=x_1^0x_2^0(mod\ 2)$ and $g=x_1x_2(mod\ 2)$. Clearly, $f\simeq_{cmr} g$, but $nof(f)=x_1^0\not\equiv x_1=nof(g)$.\qed
\end{proof}
Theorem \ref{t22} is well-illustrated by Figure \ref{f4}, in the case of   Boolean functions.

\section{Classification of Boolean functions by minor complexities}\label{sec4}
Table \ref{tb12} shows the four classes 
in $P_2^2$ under the equivalence $\simeq_{cmr}$. The $\simeq_{cmr}$-classes are represented as  union of several classes under the permuting arguments, according to Theorem \ref{t22} (ii) (see Tables \ref{tb11} and Table \ref{tb12}). 
\begin{table}[hbt]
\centering
\caption{The four  classes in  $P_2^2$ under the $cmr$-complexity.} \label{tb12}
\begin{tabular}{|ll|} \hline\hline
~[$0,\ 1$], & ~[$x_1^0x_2,\ x_1x_2^0,\ x_1\oplus x_2,\ x_1\oplus x_2^0,\ x_1^0\oplus x_1x_2,\ x_2^0\oplus x_1x_2$],\\ ~[$x_1,\ x_2,\  x_1^0,\ x_2^0$],  & ~[$x_1x_2,\ x_1\oplus x_1^0x_2,\ x_1^0\oplus x_1x_2^0,\  x_1^0x_2^0$]. \\  \hline\hline
\end{tabular} 
\end{table}

The number of  types $t(S_2^n)$ under permuting arguments,  is an upper bound of the number of equivalence classes induced by the relations $\simeq_{nof}$,  $\simeq_{cmr}$  and $\simeq_{mnr}$ for $n\le 6$ (see Table \ref{tb1}).  Theorem \ref{t22} shows that the transformation groups $NF_k^n$, $CM_k^n$ and $MN_k^n$  are not comparable with the groups  $IM_k^n$, $SB_k^n$ and $SP_k^n$,  determined by subfunction complexities.  
\begin{table}[hbt]
\centering  
\caption{Number of equivalence classes in  $P_2^n$ under transformation groups.}  \label{tb1}
\begin{tabular}{|r|rrrrrr|}\hline\hline
$n$ & $S_2^n$&$CM_2^n$&$MN_2^n$ & $IM_2^n$ &$SB_2^n$ &$SP_2^n$  \\ \hline
1&4&2&2&2&2&2\\ 
2&12&4&3&4&4&3 \\
3&80&11&5&13&11&5 \\
4&3984&*&*&104&74&11 \\
5&37 333 248&*&*&*&*& 38 \\
6&25 626 412 338 274 304&*&*&*&{*}&* \\
\hline\hline
\end{tabular}
\end{table}

\begin{figure}[hbt]
\centering
\includegraphics[trim = 0 0 0 0]{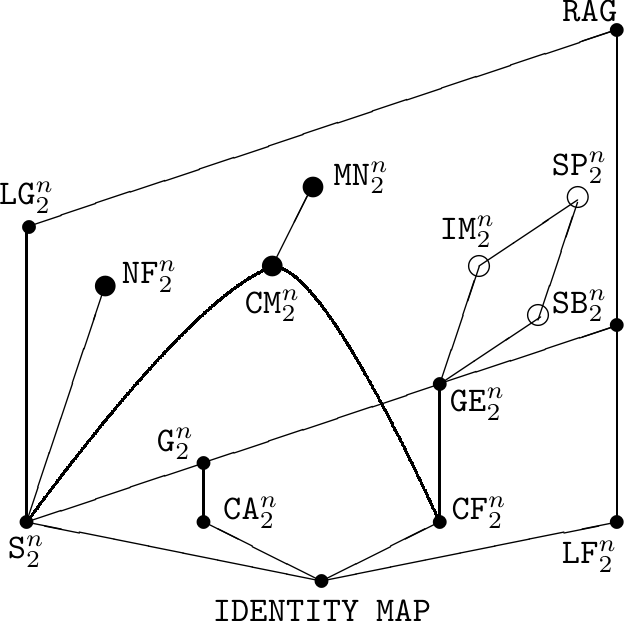}
\caption{Transformation groups in $P_2^n$.}\label{f4}

\end{figure}

Figure \ref{f4} presents the subgroups of RAG and transformation groups whose invariants are subfunction and minor complexities of Boolean functions of $n$-variables. According to Theorem \ref{t22} the group $CM_2^n$ has three subgroups from RAG, namely: $S_2^n$ - the group of permuting arguments, trivial group, consisting of the identity map, only and  $CF_2^n$- the group of complementing outputs. The groups $NF_2^n$ and  $MN_2^n$ are not subgroups of any subgroup of RAG and also, they are not subgroups of any group  among $IM_2^n$, $SP_2^n$ and $SB_2^n$. 

Table \ref{tab5} presents a full classification of the Boolean functions of tree variables by the minor complexities $cmr$ and $mnr$. If we agree to regard each $2^3$-tuple as a binary number then the last column presents the vectors of values of all ternary Boolean functions in their table representation with the natural numbers from the set $\{0,\ldots,127\}$. 
According to Theorem \ref{t16}, if a natural number $z$ presents a function $f$ which belongs to a $cmr$-class then the function $\hat f$ presented by $255-z$ belongs to the same class. Thus the catalogue (see the last column)  contents the  numbers $\le 127$, only. They represent the functions which preserve zero, i.e. the functions $f$ for which $f(0,0,0)=0$.
This classification shows that there are eleven  equivalence classes under $\simeq_{cmr}$ and five classes under $\simeq_{mnr}$.
Theorem \ref{t166} shows that each $mnr$-class is a disjoint union of  several $cmr$-classes. Thus  the first $mnr$-class consists of  all the functions which belong to the first and the second $cmr$-class (see fifth column in Table \ref{tab5}). The second $mnr$-class is equal to the third $cmr$-class. The fourth and the fifth $mnr$-classes are unions of three $cmr$-class, namely: sixth, seventh and eight, and ninth, tenth and eleventh, respectively.

\section{Appendix}\label{sec7} Table \ref{tab5} presents  classification of ternary Boolean functions under the equivalences $\simeq_{cmr}$ and $\simeq_{mnr}$, including the catalogue of the equivalence classes (last column).
\begin{example}\label{ex_last}
Let us choose a natural number belonging to  the seventh column of Table \ref{tab5}, say $24$. It belongs to the  row numbered 6. The binary representation of $24$ is $00011000$, because $24=1.2^4+1.2^3$. Hence,  the function $f$ corresponding to $24$ is evaluated by $1$ on the fourth and fifth miniterms, namely  $x_1^0x_2x_3$ and $x_1x_2^0x_3^0$. Consequently, $f=x_1^0x_2x_3\oplus x_1x_2^0x_3^0 (mod\ 2)$. Then we have $f_{2\laa 1}=f_{3\laa 1}=0$ and $f_{3\laa 2}=x_1^0x_2\oplus x_1x_2^0 (mod\ 2)$. Clearly, $cmr(f)=4$, which is written in the third cell of the sixth row. The MDD of $f$ is shown in the second cell. The $cmr$-equivalence class containing $f$ consists of 18 functions, according to the fourth cell of the sixth row and the $mnr$-equivalence class of $f$ contains 108 functions (see  whole fifth column of the table). The function $ x_1x_2x_3^0 (mod\ 2)$ is  representative for this class (sixth cell). The 
numerical list of the function from this equivalence class is given in the last seventh cell of Table \ref{tab5}.
\end{example}

\begin{center}
\begin{table}
\caption{Minor classification of ternary Boolean functions.}\label{tab5}
\begin{tabular}{|l|l|l|l|l|l|l|l|}
 \hline \hline
\multicolumn{2}{|c|}{cmr-class}&\multirow{2}{*}{cmr}&\multicolumn{2}{|c|}{Functions}&\multirow{2}{*}{mnr}&Repres.&\multirow{2}{*}{Catalogue}\\[.5ex]
 \cline{1-2} 
&MDD & & \multicolumn{2}{|c|}{per class}& &function& \\[.5ex] \hline\hline
&&&&&&&\\
1 & const &1&2&\multirow{2}{*}{}&&0&0\\ 
 \cline{1-4} \cline{7-8}
&&&&&&&\\
2& var &1&6&8&0&$x_1$&15,51,85\\
 \hline\hline
&\multirow{3}{*}{\includegraphics[trim =0 8 0 0]{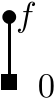}}&&&&&&\\ 
&&&&&&&10,12,34,48,60,68,\\ 
3&&2&18&18&1&$x_1x_2^0$&80,90,102\\[0ex] 
 \hline\hline
&\multirow{3}{*}{\includegraphics[trim = 0 8 0 0]{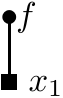}}&&&&&&\\ 
&&&&&&&3,5,17,\\ 
4&&2&12&&&$x_1x_2$&63,95,119\\[0ex] \cline{1-4}\cline{7-8}
&\multirow{3}{*}{\includegraphics[trim = 0 10 0 0]{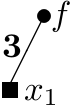}}&&&&& & \\
&&&&&&$x_1\oplus x_2$&43,77,\\  
5&&3&8&20&1&$\oplus x_3$&105,113\\[0ex] \hline\hline

&\multirow{5}{*}{\includegraphics[trim = 0 10 0 0]{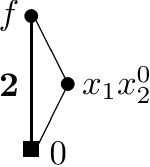}}&&&\multirow{18}{*}{}&&&\\[0ex]
&&&&&&&\\
&&&&&&& 2,4,8,\\[0ex]  
&&&&&&&16,24,32,\\[0ex] 
6&&4&18&&&$ x_1x_2x_3^0$&36,64,66\\[0ex] \cline{1-4}\cline{7-8}  


&\multirow{5}{*}{\includegraphics[trim = 0 10 0 0]{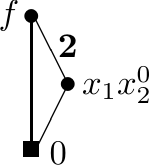}}&&&&&& \\ 
&&&&&&&6,18,20,26,28,38,\\
&&&&&&&40,44,52,56,70,72,\\
&&&&&&$x_1x_2^0x_3$&74,82,88,96,\\[.5ex]
7&&5&36&&&$\oplus x_1x_2x_3^0$&98,100\\[0ex]  \cline{1-4}\cline{7-8}

&\multirow{5}{*}{\includegraphics[trim = 0 10 0 0]{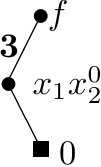}}&&&&&&14,22,30,42,46,50,\\
&&&&&&&54,58,62,76,78,84,\\ 
&&&&&&&86,92,94,104,106,\\
&&&&&&$x_1x_2^0$&108,110,112,114,116,\\
8&&6&54&108&2&$\oplus x_1x_2x_3^0$&118,120,122,124,126\\[0ex]  \hline\hline

&\multirow{5}{*}{\includegraphics[trim = 0 10 0 0]{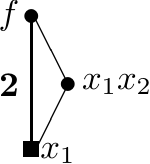}}&&&\multirow{18}{*}{}&&&7,11,13,19,21,23,\\ 
&&&&&&&31,35,41,47,49,55,\\ 
&&&&&&&59,69,73,79,81,87,\\ 
&&&&&&$x_1x_2$&93,97,107,109,115,\\
9&&4&50&&&$\oplus x_1x_2^0x_3$&117,121\\[0ex] \cline{1-4}\cline{7-8}

&&&&&&&\\

&\multirow{4}{*}{\includegraphics[trim = 0 5 0 6]{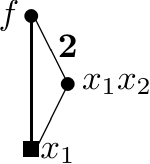}}&&&&&& 9,27,29,33,39,45,\\ 
&&&&&&&53,57,65,71,75,83,\\
&&&&&&$x_1x_2x_3$&89,99,101,\\
10&&5&36&&&$\oplus x_1x_2^0x_3^0$&111,123,125\\[0ex] \cline{1-4}\cline{7-8}


&\multirow{4}{*}{\includegraphics[trim = 0 8 0 0]{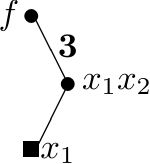}}&&&&&& \\  
 &&&&&&&1,25,37,\\ &&&&&&&61,67,91,\\
11&&6&16&102&3&$ x_1x_2x_3$&103,127\\[1ex] \hline\hline
\end{tabular}
\end{table}
\end{center}


We also provide an algorithm to find the complexity $cmr(f)$ of an arbitrary $k$-valued function $f$. Similar algorithms for manipulation of Boolean functions are presented in \cite{bra,sav}.  We shall express our algorithm in a pseudo-Pascal notation. The main data structure describes the nodes in the MDD of $f$. Each node is represented by a record declared as follows:
 
{\ttfamily type minor=record

~\quad ess: 1..n;

~\quad val: 0..$k^{k^n}-1$;

end;}

The first field named {\ttfamily ess} presents the number of essential variables in $f$ and the second field {\ttfamily val} is a natural number whose $k$-ary representation is the last column {\bf b} of the   truth table (of size $k^n$)  of $f$. For example, the function $f$ from Example \ref{ex_last} is presented as {\ttfamily f.ess=3} and {\ttfamily f.val=24}, where $k=2$ and {\bf b}= $00011000$.

The algorithm that computes \textsc{GetMinor}(g,i,j) uses well-known manipulations on the rows and columns of the truth table, which realise collapsing the $i$-th and $j$-th column and removing $m$-th row if $a_{mi}\neq a_{mj}$ for all $m=1,\ldots k^n$. This algorithm has to use a procedure, which excludes all inessential variables in the resulting minor.

The function  \textsc{GetMinor}(g,i,j) realizes one step of the reduction $\rhd$. It is also useful when constructing the MDD of a function. Then the basic algorithm in addition, has to calculate the labels of the edges in MDD and to test whether two minors are equivalent (in terms of Definition \ref{d1}).

\begin{algorithm}
  \caption{Counting cmr(f)}\label{alg1}
  \begin{algorithmic}[1]
 \State {\ttfamily type minor=record

~\quad ess: 1..n;

~\quad val: 0..$k^{k^n}-1$;

end;}
\State\texttt{var f:minor;}
\State\texttt{\quad cmr:integer;}
    \Function{GetMinor}{g:minor; i,j:integer}: minor;\Comment{Getting minor}
 \State {\ttfamily var A,H: array[1..$k^N$, 1..N] of integer (mod\ k);

~\quad B,L; array[1..$k^N$] of integer (mod\ k);

~\quad h: minor;
}    
      \State $n:=g.ess;$
             \State \texttt{Create\ truth\ table $A_{k^n\times n}B$ \ of\ $g;$}
 \State \texttt{Create\ truth\ table\ $H_{k^{n-1}\times n}L$ of\ $h:=g_{i\laa j};$}
\Comment{Use truth             
table of $g$}
        \State \texttt{Calculate\  - $h.ess$\ and\ $h.val$ from\ table\ $HL$; }        
        \State {$GetMinor:=h$};
             \EndFunction;
             
   \Function{Complexity}{g:minor}:integer;\Comment{Counting complexity}
      \State $n:=g.ess;$
      \If {$n>2$}
      \For{$j,\ 1\le j\le n-1$}
      \For{$i,\ j+1\le i\le n$}
             \State $h:=GetMinor(g,i,j);$
 \State $Complexity:=Complexity+Complexity(h);$ 
 \EndFor
 \EndFor
 \Else\Comment{Basis of recursion}
 \If{$n=2$} \State{$Complexity:=2$}\Else\State{$Complexity:=1$}
 \EndIf
 \EndIf
\EndFunction            
 \State{\texttt{Input k;\  f.ess;\ f.val;}} 
\State{\texttt{cmr:=Complexity(f).}};                    
\State{\texttt{Print\ cmr.}}
                              
  \end{algorithmic}
\end{algorithm}
\FloatBarrier

\section{Conclusion}
The traditional complexity measures  of functions are based on the reductions downto subfunctions or values of functions. They are numerical parameters of the decision diagrams built under different variable orderings. 

The minor complexities  present another concept of computing finite functions, namely when identifying variables.

 The relationship between subfunction and minor complexities in functions seems to be "strange". First of all,  
 the functions with simplest minor representations (with non-trivial arity gap) has extremely complex representations with respect to their  subfunctions (Theorem \ref{t2}). So, all functions   with at least one inseparable set  have trivial arity gap. 
 
The transformation groups whose invariants are the minor complexities have only three subgroups among the groups in RAG, namely trivial group (identity map), $S_k^n$ and $CF_k^n$, whereas the groups whose invariants are the subfunction complexities have three subgroups more, namely the groups listed  above, and $CA_k^n$, $G_k^n$, and $GE_k^n$ (see Figure \ref{f4} and Table \ref{tb2}).
 
 One of motivations to study the group $NF_k^n$ is that the reductions are inexpensive (see Algorithm \ref{alg1}, below) and the number of classes is much smaller than with (say) $GE_k^n$, because the order of $NF_k^n$ is so large.
  The order of $GE_k^n$ is $n!k^n$ (see \cite{lech1}). As mentioned, the number of equivalence classes under $NF_k^n$ equals to $k^k$. Hence,  the order of $NF_k^n$ is equal to $k^{k^n}/k^k=k^{k^{n-1}}$.
 
The most complex functions with respect to separable sets  \cite{sh5} are grouped in the largest equivalence class.  J. Denev and  I. Gyudzhenov in \cite{DG}  proved that for almost all $k$-valued functions all sets of essential variables are separable. Similar results can not be proved  for the minor complexities. For example, in $P_2^3$ the most complex   functions belong to the class numbered as 11 (see  Table \ref{tab5}), which consists of 16 function. This class is not so large. It presents 1/16 of the      all 256 ternary Boolean functions.


\nocite{}

\bibliographystyle{plain}
\bibliography{references1}
\end{document}